\documentclass[a4,12pt]{article}
\usepackage{graphics}
\usepackage[usenames]{color}
\usepackage[dvips]{epsfig}
\usepackage{mathrsfs}
\usepackage{amsmath,amssymb}
\newcommand{\be}{\begin{equation}}
\newcommand{\ee}{\end{equation}}
\newcommand{\bea}{\begin{eqnarray}}
\newcommand{\eea}{\end{eqnarray}}
\newcommand{\bean}{\begin{eqnarray*}}
\newcommand{\eean}{\end{eqnarray*}}
\newcommand{\ba}{\begin{array}}
\newcommand{\ea}{\end{array}}

\newcommand{\bc}{\begin{center}}
\newcommand{\ec}{\end{center}}
\newcommand{\bi}{\begin{itemize}}
\newcommand{\ei}{\end{itemize}}

\newcommand{\norsl}{\normalsize\sl}
\newcommand{\norsc}{\normalsize\sc}

\newcommand{\beq}{\begin{eqnarray}}
\newcommand{\eeq}{\end{eqnarray}}

\newcommand{\la}{\langle}
\newcommand{\ra}{\rangle}

\begin{document}

\title{
Drell-Yan double-spin
asymmetry $A_{LT}$ in 
polarized\\ $p \bar{p}$ collisions:
Wandzura-Wilczek contribution}   

\author{
{\norsc  Yuji Koike${}^a$, Kazuhiro Tanaka${}^b$ and Shinsuke Yoshida${}^a$}\\
\\
\norsl  ${}^a$ Department of Physics, Niigata University,
Ikarashi, Niigata 950-2181, Japan 
\\
\norsl  ${}^b$ Department of Physics, Juntendo University,
    Inba, Chiba 270-1695, Japan}
\date{}
\maketitle

\begin{abstract}
The longitudinal-transverse 
spin asymmetry $A_{LT}$ in the polarized
Drell-Yan process depends on the twist-3 spin-dependent distributions of nucleon. 
In addition to 
the contributions expressed as
matrix element of
the twist-3
operators,
these distributions contain
the so-called Wandzura-Wilczek part,
which 
is completely determined by 
a certain integral of 
the twist-2 spin-dependent parton distributions.
We demonstrate that the recently obtained empirical information 
on the transversity distribution
allows a realistic estimate of the Wandzura-Wilczek contribution to $A_{LT}$ 
for the case of 
polarized proton-antiproton
collisions.
In particular, our results in the Wandzura-Wilczek approximation
indicate that rather large $A_{LT}$ can be observed in the proposed 
spin experiments at GSI, and its 
behavior as a function of dilepton mass
obeys novel pattern, 
compared with the other double-spin asymmetries 
$A_{TT}$ and $A_{LL}$.
Our results provide a guide for testing 
a signal of
effects originating from the twist-3 operators
associated with quark-gluon correlation.
\end{abstract}

\newpage

\setcounter{equation}{0}
\renewcommand{\theequation}{\arabic{equation}}

\vskip 1cm
The proposed polarization experiments 
with antiprotons at GSI \cite{PAX:05} 
stimulate renewed interest in the polarized Drell-Yan processes
to access the chiral-odd spin-dependent parton distributions of the nucleon.
The double transverse-spin asymmetries $A_{TT}$
for lepton pair production 
in collisions of transversely polarized protons and antiprotons,
$p^{\uparrow}\bar{p}^{\uparrow}\rightarrow l^+l^-X$,
are estimated, 
and are found to be large enough to be measured 
at GSI \cite{Anselmino:2004ki}, providing promising way to probe
the chiral-odd twist-2 spin-dependent parton distribution,
the transversity $h_1 (x)$~\cite{RS,JJ,CPR:92,BDR:02,KT:99}.
In particular, the $p\bar{p}$ collisions at moderate energy in GSI experiments
allow us to probe the relevant parton distributions in the ``valence region'',
in contrast to 
the complementary case of $pp$ collisions at, e.g., RHIC
where 
the ``sea-quark region'' is mainly probed.
The QCD corrections to $A_{TT}$ at GSI have been studied recently
at next-to-leading order (NLO)~\cite{BCCGR:06} 
and at higher orders with the ``threshold resummation''~\cite{SSVY:05}.
The resummation corrections relevant when the transverse-momentum
of the produced lepton pair is small~\cite{KKST:06,KKT:07-2} are also investigated~\cite{KKT:08}.
It has been found that 
the behavior of these QCD corrections associated with the valence region 
is rather different from the corresponding effects involving the sea quarks
for the $pp$-collision cases~\cite{MSSV:98,KKT:07,KKT:07-2}.
As a result,
these QCD corrections are
small 
at the kinematical regions corresponding to the GSI experiments,
suggesting that the large LO $A_{TT}$ at GSI
is rather robust.
This fact also allows us to estimate the value of $A_{TT}$ at GSI
using only the 
empirical information on the transversity 
distributions~\cite{KKT:08}, 
which is recently extracted~\cite{Anselmino:07} 
through the LO global fit to the semi-inclusive deep inelastic scattering (SIDIS) data, 
in combination 
with the $e^{+}e^{-}$ data for the associated (Collins) fragmentation function.

\smallskip

The PAX Collaboration has proposed 
the Drell-Yan experiments
in $p\bar{p}$ collisions at the CM energy $\sqrt{s}$ with  $s=30$ and $45$~GeV$^2$ 
in the fixed-target mode, 
and those up to $s= 210$~GeV$^2$ in the collider mode~\cite{PAX:05}.
Those GSI-PAX experiments will measure $A_{TT}$ for $0.2 \lesssim Q/\sqrt{s}\lesssim 0.7$
with $Q$ the mass of the produced dilepton, and indeed probe
the transversity $h_1 (x)$ in the valence region
in a wide range of $x$. 
It should not be overlooked that 
the double-spin 
longitudinal-transverse asymmetry $A_{LT}$ is also readily
accessible in those Drell-Yan experiments, in particular, in the fixed-target mode
with the longitudinal polarization of the target:
$A_{LT}$ plays a distinguished role in spin physics
because it allows us to access the twist-3 spin-dependent parton distributions 
as leading effects~\cite{JJ}, 
similarly as the longitudinal-transverse asymmetry associated with the 
structure function $g_2$ in the polarized DIS~\cite{g2exp}.
Thus the data of $A_{LT}$ will provide an experimental test whether the quark-gluon-quark 
correlations inside the nucleon is sizeable or not, in particular, in 
the chiral-odd spin structure that is  
not accessible by $g_2$ in DIS\footnote{For a test of chiral-odd quark-gluon-quark 
correlation using the SIDIS data, see \cite{Avakian:2007mv}.}.
These facts call for theoretical study of $A_{LT}$ to assess its potential at GSI experiments,
which is the purpose of this Letter.
Up to now $A_{LT}$ was estimated
for the $pp$-collision cases~\cite{Kanazawa:1998rw},
but the above mentioned situation for $A_{TT}$ suggests that the behavior 
of $A_{LT}$ will be different between the $pp$ and $p\bar{p}$ collisions.
Also it is important to clarify the impact of the 
new empirical information~\cite{Anselmino:07,Prokudin} of the transversity 
distribution $h_1 (x)$ on the prediction of $A_{LT}$ at GSI,
because $A_{LT}$ depends on $h_1 (x)$. 
We will demonstrate that
this empirical information for $h_1 (x)$
indeed allows a useful estimate for $A_{LT}$ 
in $p\bar{p}$ collisions at GSI kinematics.

\smallskip

We shall work at LO QCD, which provides a sufficient accuracy
for our first estimate of $A_{LT}$ at GSI. We may anticipate that
the mechanism associated with the valence region relevant to GSI kinematics
could make the QCD corrections to $A_{LT}$ small, similarly to the case for $A_{TT}$
mentioned above.
To calculate $A_{LT}$, we first recall the parton distributions of the nucleon.
At LO, we need the spin-dependent quark distribution
functions of twist-3 as well as of twist-2,
which are defined as the nucleon matrix element of the chiral-odd and chiral-even
quark bilocal operator with the light-like separation 
between the constituent fields~\cite{JJ,BDR:02,KT:99},
\beq
& &\int{d\lambda \over 2\pi} e^{i\lambda x}\la PS|\bar{\psi}
(0)\sigma_{\mu\nu} i\gamma_5 \psi(\lambda n) |PS \ra 
\nonumber\\
& & \qquad=
2\left[ h_1(x,\mu^2)\left( S_{\perp\mu}P_\nu -  S_{\perp\nu}P_\mu \right)/M
+h_L(x,\mu^2)M \left( P_\mu n_\nu - P_\nu n_\mu \right) (S\cdot n)
\right],
\label{eq1}\\
& &\int{d\lambda \over 2\pi} e^{i\lambda x}\la PS|\bar{\psi}
(0)\gamma_\mu \gamma_5 \psi( \lambda n ) |PS \ra  
=2\left[ g_1(x,\mu^2)P_\mu (S\cdot n) + g_T(x,\mu^2) 
S_{\perp\mu} \right],
\label{eq2}
\eeq
and we also need the unpolarized quark distribution defined as usually as
\be
\int{d\lambda \over 2\pi} e^{i\lambda x}\la PS|\bar{\psi}
(0)\gamma_\mu \psi( \lambda n ) |PS \ra  = 2 f_1(x,\mu^2)P_\mu,
\label{eq3}
\ee
where $|PS\ra$ denotes the nucleon state with mass $M$, the four momentum
$P^\mu=(P^+, P^-, \bold{0}_{\perp})$, and the spin vector 
$S^\mu$ 
satisfying $P^2 = 2P^+ P^- =M^2$, $S^2=-M^2$, and $P\cdot S =0$,
and a light-like vector $n^\mu=(0, n^-, \bold{0}_{\perp})$  
is introduced by the relation
$P\cdot n=1$.
$S^\mu$ is decomposed as $S^\mu=(S\cdot n) P^\mu -M^2 (S\cdot n) n^\mu + 
S_\perp^\mu$ with $P\cdot S_\perp= n\cdot S_\perp=0$.  
In (\ref{eq1})-(\ref{eq3}), the gauge link operators which ensure gauge invariance
are suppressed for simplicity.
The distribution functions $h_{1,L}$, $g_{1,T}$ and $f_1$
depend on the 
factorization scale $\mu$, at which the bilocal operators in the LHS are renormalized,
and those distribution functions 
are defined for each quark and anti-quark flavor
$\psi = \psi^a$ ($a=u, \bar{u}, d, \bar{d}, s, \bar{s},\ldots$) as $h_{1,L}^a$, etc.
We remind that in the infinite momentum frame ($P^+\to \infty$) the
Lorentz structures associated with $h_1$, $g_1$ and $f_1$ are of $O(P^+)$ (twist-2),
those for $h_L$ and $g_T$ are of $O(1)$ (twist-3), and those behaving as twist-4
($O(1/P^+)$) 
are ignored
in the RHS of (\ref{eq1})-(\ref{eq3}).
$g_1^u(x , \mu^2)$ is the familiar helicity distribution for $u$-quark 
carrying the momentum component $k^+=xP^+$
inside the longitudinally polarized nucleon, and, similarly, $h_1^a(x, \mu^2)$ 
is the transversity distribution
inside the transversely polarized nucleon~\cite{RS,JJ}.
Note that the twist-3 distributions $h_L$ and $g_T$ are also associated with
the longitudinal and transverse polarization of the nucleon, respectively.

\smallskip

The above mentioned classification of twist based on the power counting
in the infinite momentum frame is directly related to the power of $1/Q$
with which the corresponding distributions appear in the physical cross sections,
but does not exactly match the conventional and formal definition of twist as 
``dimension minus spin'' associated with the relevant operator structure
in (\ref{eq1}) and (\ref{eq2}).
As a result, the distributions $h_L$ and $g_T$ 
actually contain the piece that is expressed by matrix element of the twist-2 operators
as~\cite{WW:77,JJ,KT:99}
(see also Appendix in \cite{EKT:06})
\beq
h_L^a(x,\mu^2) &=& 2x\int_{x}^{1}dy \frac{h_1^a(y,\mu^2)}{y^2}  
+\cdots,
\label{hLWW}\\
g_T^a(x,\mu^2) &=& \int_{x}^{1} dy\frac{g_1^a(y,\mu^2)}{y}
+\cdots,
\label{gTWW}
\eeq
where the ellipses stand for ``genuine twist-3'' contributions 
given as matrix element of the twist-3 operators; it is known that those
twist-3 operators can be reexpressed as 
quark-gluon-quark three-body correlation operators on the lightcone,
using the QCD equations of motion~\cite{JJ,g2,KT,BBKT}.  
In the following we call the twist-2 component, 
shown explicitly in (\ref{hLWW}) and (\ref{gTWW}), 
the Wandzura-Wilczek part. 
Because the operators with different geometric twist 
do not mix with each other under renormalization,   
the Wandzura-Wilczek part
does not mix with the genuine twist-3 contributions under the QCD evolution with $\mu^2$.
Thus both $x$- and $\mu^2$-dependences of the Wandzura-Wilczek part
are determined solely by those of the twist-2 distribution functions 
as (\ref{hLWW}) and (\ref{gTWW}).
Taking into account only the Wandzura-Wilczek part  
in (\ref{hLWW}) and (\ref{gTWW}) yields the ``Wandzura-Wilczek approximation'' for 
$h_L$ and $g_T$.

\smallskip
 
With the above definitions for the parton distributions, 
we can write down the LO expression for the longitudinal-transverse spin asymmetry $A_{LT}$
in the polarized $p\bar{p}$ collisions. Before doing this, it is worthwhile to remind 
the LO formula of the other double-spin asymmetries~$A_{LL}$ and $A_{TT}$~\cite{RS,JJ,BDR:02}. 
Using the quark distributions inside the proton,
\beq
A_{LL} &=& 
\frac{\frac{d\sigma^{\rightarrow \rightarrow}}{dQ^2 dx_F d\Omega} 
-\frac{d\sigma^{\rightarrow\leftarrow}}{dQ^2 dx_F d\Omega}}
{\frac{d\sigma^{\rightarrow \rightarrow}}{dQ^2 dx_F d\Omega} 
+\frac{d\sigma^{\rightarrow\leftarrow}}{dQ^2 dx_F d\Omega}}               
       =  \hat{a}_{LL}
\frac{\sum_{a} e_{a}^2
                   g_{1}^a(x_1,Q^2)g_{1}^{a}(x_2,Q^2)}
               {\sum_{a} e_{a}^2 f_{1}^a(x_1,Q^2)f_{1}^{a}(x_2,Q^2)},
\label{ALL}\\[5pt]
A_{TT} &=& 
\frac{\frac{d\sigma^{\uparrow \uparrow}}{dQ^2 dx_F d\Omega} 
-\frac{d\sigma^{\uparrow\downarrow}}{dQ^2 dx_F d\Omega}}
{\frac{d\sigma^{\uparrow \uparrow}}{dQ^2 dx_F d\Omega} 
+\frac{d\sigma^{\uparrow\downarrow}}{dQ^2 dx_F d\Omega}}              
       = \hat{a}_{TT}  \frac{\sum_{a} e_{a}^2
                   h_{1}^a(x_1,Q^2)h_{1}^{a}(x_2,Q^2)}
               {\sum_{a} e_{a}^2 f_{1}^a(x_1,Q^2)f_{1}^{a}(x_2,Q^2)},
\label{ATT}
\end{eqnarray}
for the production of the dilepton with the invariant mass $Q$ and the 
longitudinal momentum component $Q_z$ corresponding to the Feynman $x_F$,
where one of the leptons outgoes 
to the direction with the angle
$\Omega = (\theta, \phi)$.
$e_a$ represents the electric charge of the quark-flavor
$a$ and the summation is over all quark and anti-quark flavors, 
$a=u, \bar{u}, d, \bar{d}, s, \bar{s},\ldots$. 
The scaling variables $x_{1,2}$ 
represent the momentum fractions associated with the partons
annihilating via the Drell-Yan mechanism, such that 
$Q^2=(x_1 P_1 + x_2 P_2)^2 = x_1 x_2 s$ and $x_F=x_1-x_2$ ($=2Q_z/\sqrt{s}$ in the CM
frame), where $s=(P_1 +P_2)^2$ is
the CM energy squared of the colliding proton and antiproton.
This implies
\beq
x_1 = {1\over 2}\left( x_F +\sqrt{x_F^2 + {4Q^2\over s}}\right), \qquad
x_2 = {1\over 2}\left( -x_F +\sqrt{x_F^2 + {4Q^2\over s}}\right).
\label{xf}
\eeq
In (\ref{ALL}) and (\ref{ATT}), 
$\hat{a}_{LL}$ and $\hat{a}_{TT}$ represent
the asymmetries in the parton level defined as
\be
\hat{a}_{LL}  = 1,  
 \qquad\qquad\hat{a}_{TT}  =  \frac{{\rm sin}^2\theta\, 
{\rm cos}2\phi}{1+{\rm cos}^2\theta},
\label{aTT}
\ee
with the polar and azimuthal angles $\theta$ and $\phi$ 
in the dilepton rest frame
with respect to the incoming beam and transverse-spin axes,
respectively.
The LO formula
for $A_{LT}$ in $p^{\rightarrow}\bar{p}^{\uparrow}\rightarrow l^+l^-X$ or
$p^{\uparrow}\bar{p}^{\rightarrow}\rightarrow l^+l^-X$
can be expressed similarly as~\cite{JJ,BDR:02}
\be
A_{LT} = 
\frac{\frac{d\sigma^{\rightarrow \uparrow}}{dQ^2 dx_F d\Omega} 
-\frac{d\sigma^{\rightarrow\downarrow}}{dQ^2 dx_F d\Omega}}
{\frac{d\sigma^{\rightarrow \uparrow}}{dQ^2 dx_F d\Omega} 
+\frac{d\sigma^{\rightarrow\downarrow}}{dQ^2 dx_F d\Omega}}
       = \hat{a}_{LT} \frac{\sum_{a} e_{a}^2
                   \left [g_{1}^a(x_1,Q^2)x_2g_{T}^{a}(x_2,Q^2)
                        + x_1h_{L}^a(x_1,Q^2)h_{1}^{a}(x_2,Q^2) \right]}
               {\sum_{a} e_{a}^2 f_{1}^a(x_1,Q^2)f_{1}^{a}(x_2,Q^2)},       
\label{ALT}
\ee
associating
the variables $x_1$ and $x_2$ with the longitudinally and 
transversely polarized beams, respectively, with
\be
\hat{a}_{LT}  =  \frac{M}{Q} \frac{2\,{\rm sin}2\theta\, 
{\rm cos}\phi}{1+{\rm cos}^2\theta}.
\label{aLT}
\ee
We note that $A_{LL}$ and $A_{TT}$ receive contribution only from
the twist-2 distributions, while $A_{LT}$ is proportional to the twist-3 
distributions
and hence $\hat{a}_{LT}$ is suppressed by a factor $1/Q$ compared with (\ref{aTT}).

\smallskip

To compute 
the above formulae (\ref{ALL}), (\ref{ATT}), and (\ref{ALT}) 
with the GSI kinematics, 
we have to specify the LO parton distributions to be substituted. 
We use the LO GRV98~\cite{GRV:98} 
and GRSV2000 (``standard scenario'')~\cite{GRSV:00} distributions  
for the unpolarized and longitudinally-polarized quark distributions 
$f_1^a(x,Q^2)$ and $g_1^a(x,Q^2)$, respectively.
For the LO transversity distribution $h_1^a(x,Q^2)$,
we are guided by the recent information from the LO global fit~\cite{Anselmino:07,Prokudin}:
we find that a useful estimate can be obtained
by assuming the relation 
\bea
h_1^a (x,\mu^2)=g_1^a(x,\mu^2),
\label{h1g1}
\eea
at a low scale $\mu$ ($\mu^2=0.26$ GeV$^2$ using the GRSV2000 $g_1^a(x,\mu^2)$);
its QCD evolution from $\mu^2$ to $Q^2$ is controlled by 
the LO DGLAP kernel~\cite{KMHKKV:97} for the transversity.
It is worth noting that the above relation (\ref{h1g1}) at the low $\mu^2$,
which is exact in the non-relativistic limit,
is suggested also by the estimates 
from relativistic quark models for nucleon~\cite{BDR:02,Anselmino:2004ki,waka},  
matches the results by lattice QCD simulation~\cite{SFQCD:05,RBRC},
and has been used 
in the previous estimates for $A_{TT}$ 
at GSI~\cite{Anselmino:2004ki,BCCGR:06,KKT:08}.
%
\begin{figure}
\bc
\includegraphics[height=7cm]{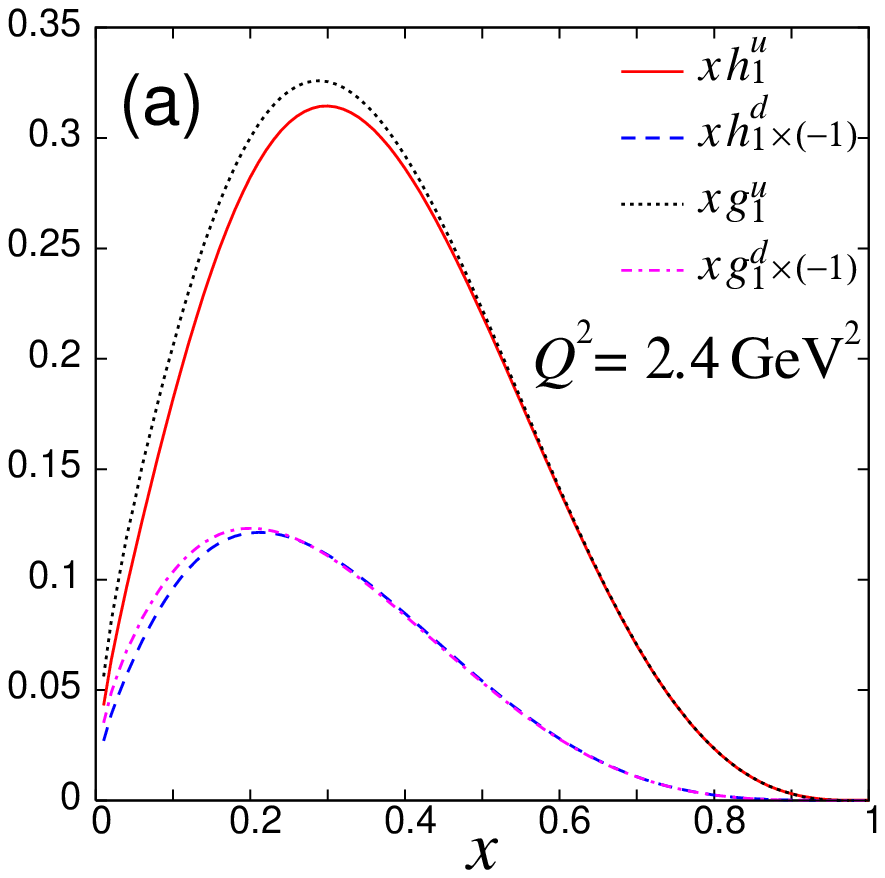}~~
\includegraphics[height=7cm]{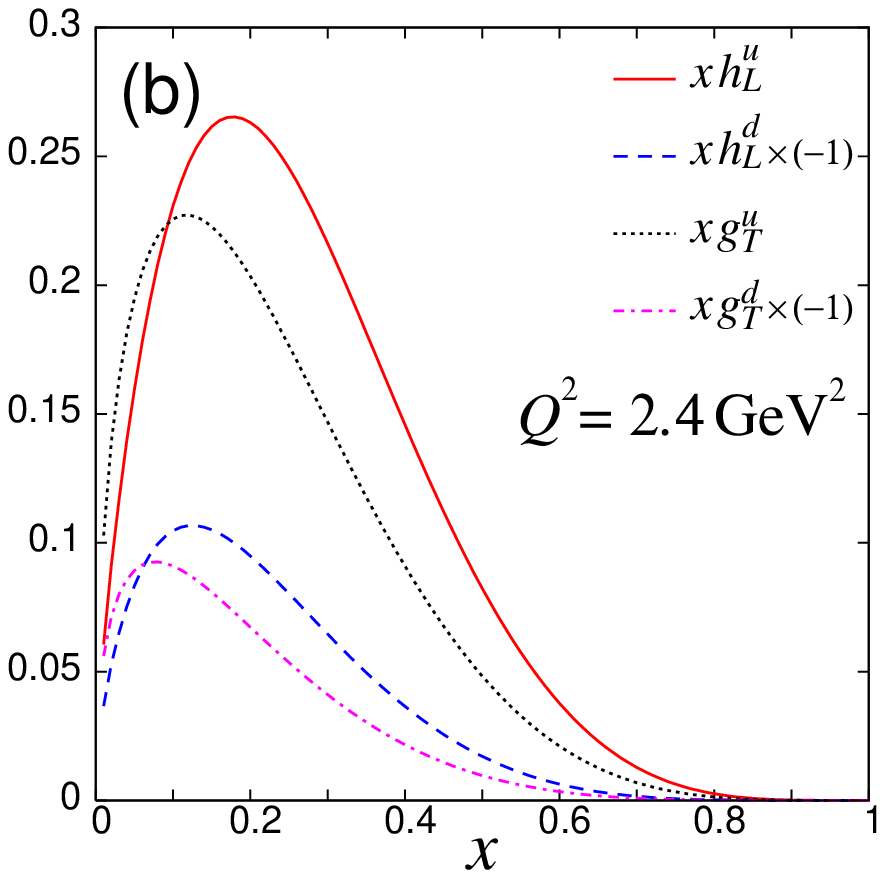}
\ec
\caption{(a)~The transversity distributions $xh_1^a(x,Q^2)$ 
and 
the helicity distributions $xg_1^a(x,Q^2)$ at the scale $Q^2=2.4$ GeV$^2$
for $u$- and $d$-quarks.  (b)~The twist-3 distributions $xh_L^a(x,Q^2)$ and $xg_T^a(x,Q^2)$ 
in the Wandzura-Wilczek approximation
for $u$- and $d$-quarks.
For convenience, we multiplied $-1$ for $d$-quark distributions in both figures. 
}
\label{fig:1}
\end{figure}
%
The obtained LO transversity distributions for $u$ and $d$ quarks, 
$x h_1^u (x,Q^2)$ and $x h_1^d (x,Q^2)$,
are shown in Fig.~\ref{fig:1}(a) as a function of $x$
at $Q^2 =2.4$~GeV$^2$. 
$xg_1^{u,d} (x,Q^2)$ is also shown in the same figure.
For convenience, 
we have 
multiplied the factor $-1$ to the $d$-quark distributions.
If we compare Fig.~\ref{fig:1}(a) with the results of the LO global 
fit~\cite{Anselmino:07,Prokudin},
we see that, for the valence region $0.2 \lesssim x \lesssim 0.7$ relevant 
for the GSI kinematics, 
our LO transversities lie slightly outside the error band of the fit,
similarly as observed for the NLO case~\cite{KKT:08}.
Therefore, our transversities will provide a realistic estimate of  
the upper bound of the relevant asymmetries, implied by the present empirical uncertainty 
in the transversities. (At present there are no data to constrain the 
transversity $h_1^a (x,Q^2)$ directly for $x > 0.4$, and in this region
the uncertainty bands resulting from the LO global fit~\cite{Anselmino:07,Prokudin}  
could be subject to the particular choice of the parameterization of $h_1^a (x,Q^2)$ 
assumed in the fitting procedure.)
We see from 
Fig.~\ref{fig:1}(a) that
$\left( h_1^u (x,Q^2)\right)^2 \gg \left( h_1^d (x,Q^2)\right)^2$ 
and $\left( g_1^u (x,Q^2)\right)^2 \gg \left( g_1^d (x,Q^2)\right)^2$ in the valence region, 
and likewise~\cite{GRV:98} for 
$f_1^a(x,Q^2)$. 
This also holds
for higher $Q^2$,
so that 
\be
\frac{A_{TT}}{ \hat{a}_{TT}} 
       \simeq \frac{h_{1}^u(x_1,Q^2)h_{1}^{u}(x_2,Q^2)}
               {f_{1}^u(x_1,Q^2)f_{1}^{u}(x_2,Q^2)},
\label{ATT2}
\ee
for (\ref{ATT}) at GSI, and likewise for (\ref{ALL}).
Hence the GSI measurement of (\ref{ATT}) allows a direct access to 
$h_1^u (x,Q^2)$~\cite{Anselmino:2004ki,BCCGR:06,KKT:08}\footnote{
For the values of $Q \gtrsim 1$ GeV relevant to the Drell-Yan process, 
our LO transversities satisfy Soffer's inequality~\cite{Soffer:95},
$2 |h_1^a (x, Q^2)| \le f_1^a(x,Q^2)+ g_1^a(x,Q^2)$, 
for the $u$-, $\bar{u}$- and $\bar{d}$-quarks,
but violate it for the $d$-,
$s$- and $\bar{s}$-quarks
by a small amount, similarly as in the previous works~\cite{KKT:08,Kanazawa:1998rw},
because of negative polarization for the $d$-quark, $g_1^d(x,Q^2) <0$, 
and the smallness of $s$- and $\bar{s}$-densities 
($f_1^s(x,\mu^2)=f_1^{\bar{s}}(x,\mu^2)=0$ at the input scale $\mu$ in GRV98).
This violation of Soffer's inequality will be harmless to our
numerical estimates of the asymmetries because of 
the dominance of the $u$-quark distribution as (\ref{ATT2}) and (\ref{ALT2}).}.

\smallskip

The twist-2 spin-dependent distributions obtained above
determine $h_L$ and $g_T$ 
in the Wandzura-Wilczek approximation
using (\ref{hLWW}) and (\ref{gTWW}).
The results for the $u$- and $d$-quarks with $Q^2 =2.4$ GeV$^2$ are shown in Fig.~\ref{fig:1}(b),
with the curves for the $d$ quark 
showing the results multiplied by $-1$.
The integral with the factors $1/y$, $1/y^2$ 
in (\ref{hLWW}), (\ref{gTWW}) shifts the peak of the curves 
to lower $x$ with the suppressed peak-height, compared with those
for the corresponding twist-2 distributions in Fig.~\ref{fig:1}(a).  
We also see the $u$-quark dominance in the valence region,
similarly as in the twist-2 distributions, so that
\be
\frac{A_{LT}}{\hat{a}_{LT}} \simeq
\frac{g_{1}^u(x_1,Q^2)x_2\int_{x_2}^{1} dy\frac{g_1^u(y,Q^2)}{y}
 + 2x_1^2 \int_{x_1}^{1}dy \frac{h_1^u(y,Q^2)}{y^2}  h_{1}^{u}(x_2,Q^2)}
               {f_{1}^u(x_1,Q^2)f_{1}^u(x_2,Q^2)} + \cdots,
\label{ALT2}
\ee
for (\ref{ALT}) at GSI kinematics, where the ellipses denote the 
contributions associated with the genuine twist-3 operators.
This implies that the ``Wandzura-Wilczek contribution'' to $A_{LT}$ at GSI is
directly related to the behavior of $h_1^u (x,Q^2)$. 

\smallskip

The data on the transverse spin structure function $g_2$ from the polarized DIS experiments 
indicate
that the genuine twist-3 contribution in (\ref{gTWW}) is small\,\cite{g2exp},
and $g_T$ approximately follows the Wandzura-Wilczek result.
The calculations of low moments of $g_T$ by lattice QCD simulation 
support this result\,\cite{RBRC,lattice-g2}.
Also, estimates from nucleon models, combined with the QCD evolution for the relevant
twist-3 operators~\cite{ABH,BBKT}, suggest 
that the Wandzura-Wilczek part of (\ref{gTWW}) and (\ref{hLWW}) 
dominates $g_T$ and $h_L$ for $\mu^2 \gg 1$ GeV$^2$~\cite{KK,St,Kanazawa:1998rw}.
For the present first estimate of $A_{LT}$
of (\ref{ALT}) in $p\bar{p}$ collisions, we employ the Wandzura-Wilczek approximation
of Fig.~\ref{fig:1}(b) for $g_T$ and $h_L$.

\begin{figure}
\bc
\includegraphics[height=7cm]{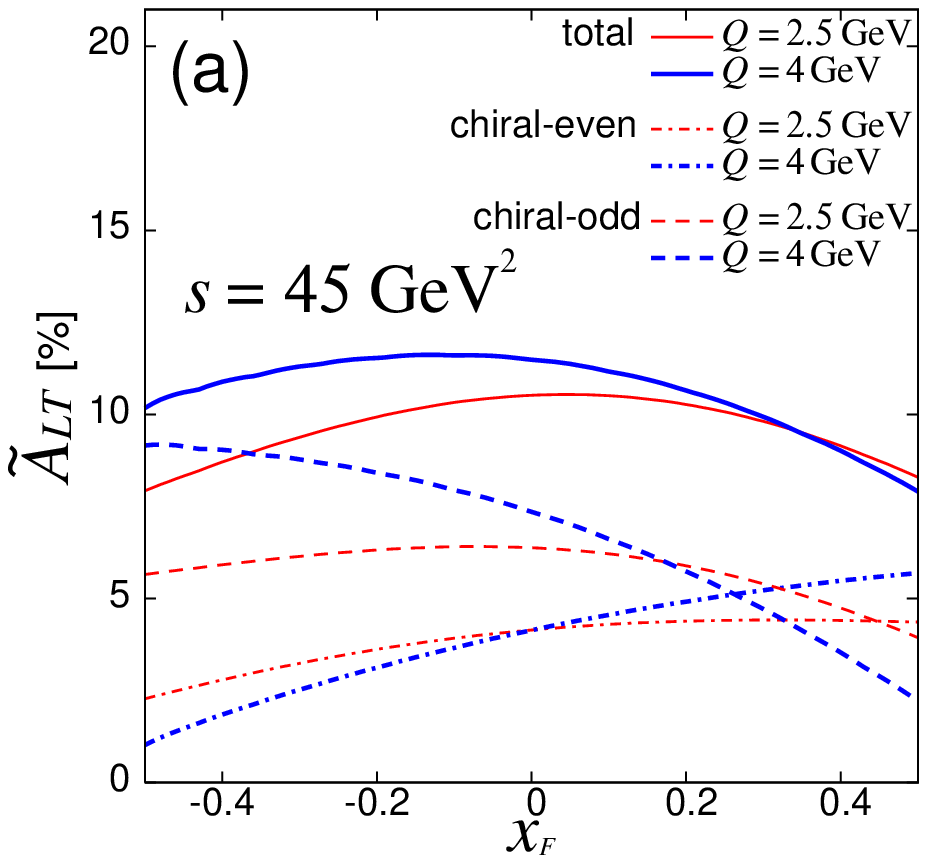}~~
\includegraphics[height=7cm]{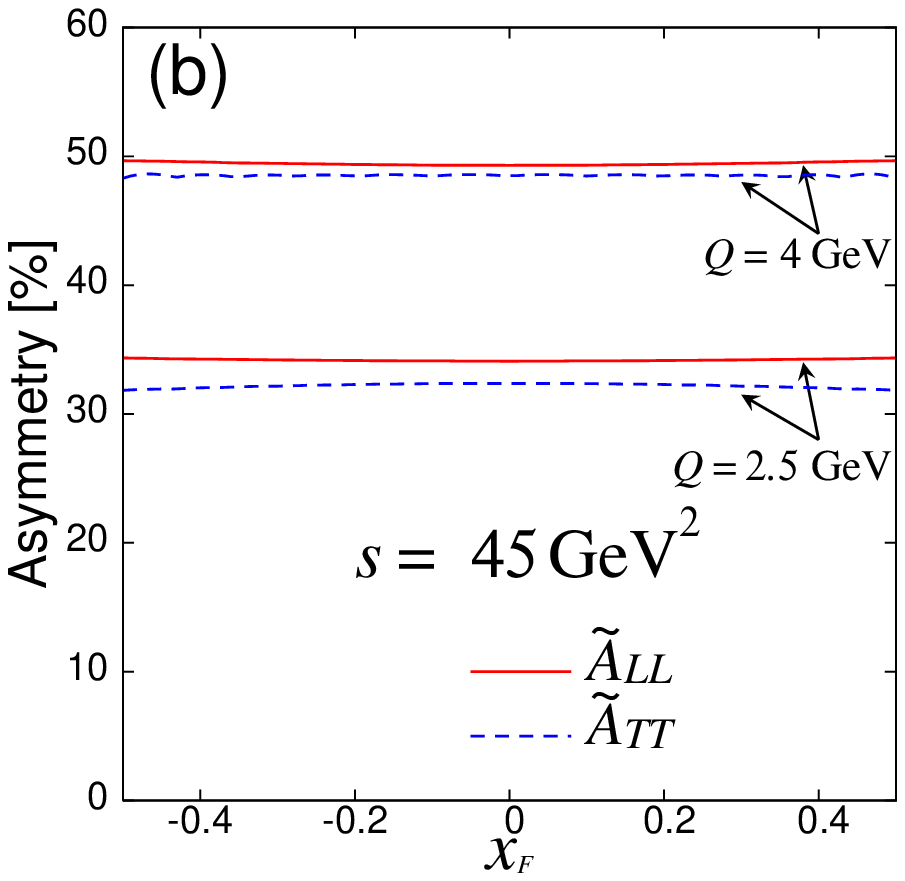}
\ec
\caption{(a)~$\widetilde{A}_{LT}$ and (b)~$\widetilde{A}_{LL}$ and $\widetilde{A}_{TT}$ 
as a function of $x_F$ for $Q=2.5$ and $4$ GeV at $s=45$ GeV$^2$.
For $\widetilde{A}_{LT}$, the chiral-even and -odd contributions are also
shown separately.  
}
\label{fig:2}
\end{figure}
%

\smallskip

In all the following numerical evaluation, we present the 
results for the ``reduced asymmetries'' $\widetilde{A}_{YW}\equiv A_{YW}/\hat{a}_{YW}$
($Y,W = L, T$)\footnote{Note that
$\hat{a}_{LT}$ of (\ref{aLT}) is defined absorbing the suppression factor $M/Q$
specific to twist-3 cross section.}. 
We first consider the fixed-target mode,
where $A_{LT}$ will be readily accessible.
Figure \ref{fig:2}(a) shows $\widetilde{A}_{LT}$ 
as a function of $x_F$ for $Q=2.5$ and $4$ GeV at $s=45$ GeV$^2$.
Also shown are the separated contributions from the chiral-even and -odd
distributions, corresponding to the first and second terms in the numerator in (\ref{ALT}).
The results may be compared with the behavior of $\widetilde{A}_{LL}$ and $\widetilde{A}_{TT}$ 
at the same kinematics, shown in 
Fig.~\ref{fig:2}(b). The curves for 
$\widetilde{A}_{TT}$ reproduce the corresponding LO results in \cite{BCCGR:06}.
$\widetilde{A}_{LL}$ and 
$\widetilde{A}_{TT}$ are symmetric with respect to $x_F=0$, while
$\widetilde{A}_{LT}$ is not symmetric (compare (\ref{ALL}), (\ref{ATT}) with (\ref{ALT})).
$\widetilde{A}_{LL}$ and 
$\widetilde{A}_{TT}$ are almost flat as a function of $x_F$ for the GSI kinematics~\cite{BCCGR:06},
in strong contrast to $\widetilde{A}_{LT}$. These features of $\widetilde{A}_{LT}$ 
come from the $x_F$ dependence of
chiral-even and -odd contributions; in particular, the chiral-odd contribution 
shows the tendency to increase for decreasing $x_F$, while the chiral-even one shows
opposite tendency.
The values of $\widetilde{A}_{LL}$ and $\widetilde{A}_{TT}$ 
are more than 30\% and are much larger than their typical values 
in the $pp$-collision cases~\cite{MSSV:98,Kanazawa:1998rw}. 
This is because of the fact that for the GSI kinematics the valence contributions are dominant 
both in the numerator and the denominator of
(\ref{ALL}) and (\ref{ATT})\footnote{$\widetilde{A}_{LL}$ is slightly larger 
than $\widetilde{A}_{TT}$, because $g_1^u$ is slightly larger than $h_1^u$
as in Fig.~\ref{fig:1}(a).}
and the small-$x$ rise
of sea-distributions is absent in the denominator~\cite{Anselmino:2004ki,SSVY:05,BCCGR:06,KKT:08}.
We see in Fig.~\ref{fig:2}(a)
that the similar mechanism leads to the significant value ($\gtrsim 10$\%)
also for $\widetilde{A}_{LT}$.
In general, $\widetilde{A}_{LT}$ is smaller than $\widetilde{A}_{LL}$, $\widetilde{A}_{TT}$ 
by the presence of the additional factor, $x_1$ or $x_2$, in (\ref{ALT}) compared with 
(\ref{ALL}), (\ref{ATT}). 
Further suppression effect for $\widetilde{A}_{LT}$ could be caused by the behavior of 
$g_T$ and $h_L$
observed in Fig.~\ref{fig:1}(b) in comparison with Fig.~\ref{fig:1}(a). 
When the sea-quark region is probed in $pp$ collisions,
these effects, in particular the additional $x_{1,2}$ factor, 
lead to $\widetilde{A}_{LT}$ much smaller than the corresponding 
$\widetilde{A}_{LL}$, $\widetilde{A}_{TT}$, as demonstrated in \cite{Kanazawa:1998rw}.

\begin{figure}
\bc
\includegraphics[height=7cm]{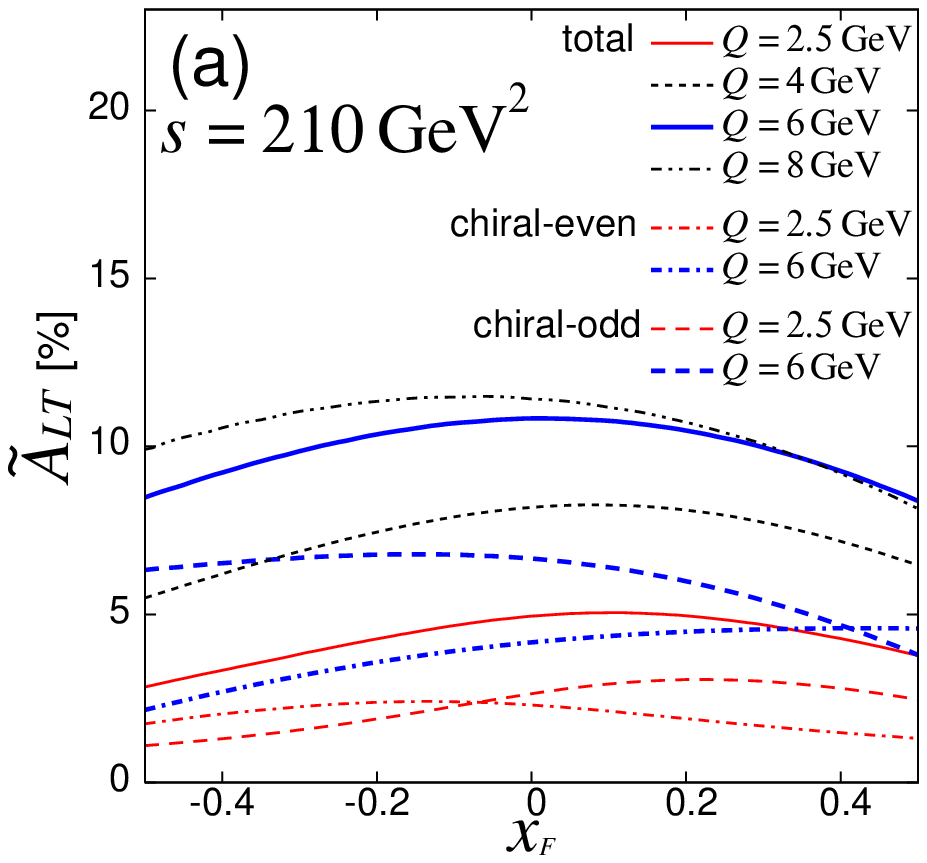}~~
\includegraphics[height=7cm]{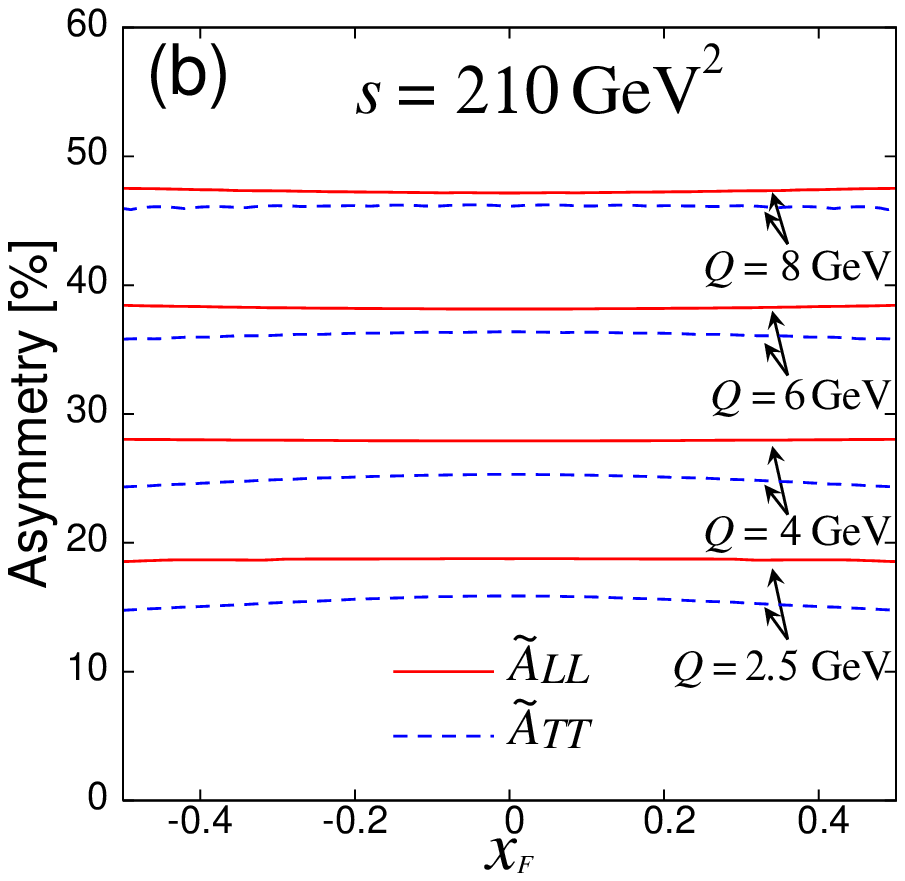}
\ec
\caption{(a)~$\widetilde{A}_{LT}$ and (b)~$\widetilde{A}_{LL}$ and $\widetilde{A}_{TT}$ 
as a function of $x_F$ for $Q=2.5$, $4$, $6$ and $8$ GeV at $s=210$ GeV$^2$.
For $\widetilde{A}_{LT}$ with $Q=2.5$ and $6$ GeV, the chiral-even and -odd contributions are also
shown separately.  
}
\label{fig:3}
\end{figure}

\smallskip

\begin{figure}
\bc
\includegraphics[height=7cm]{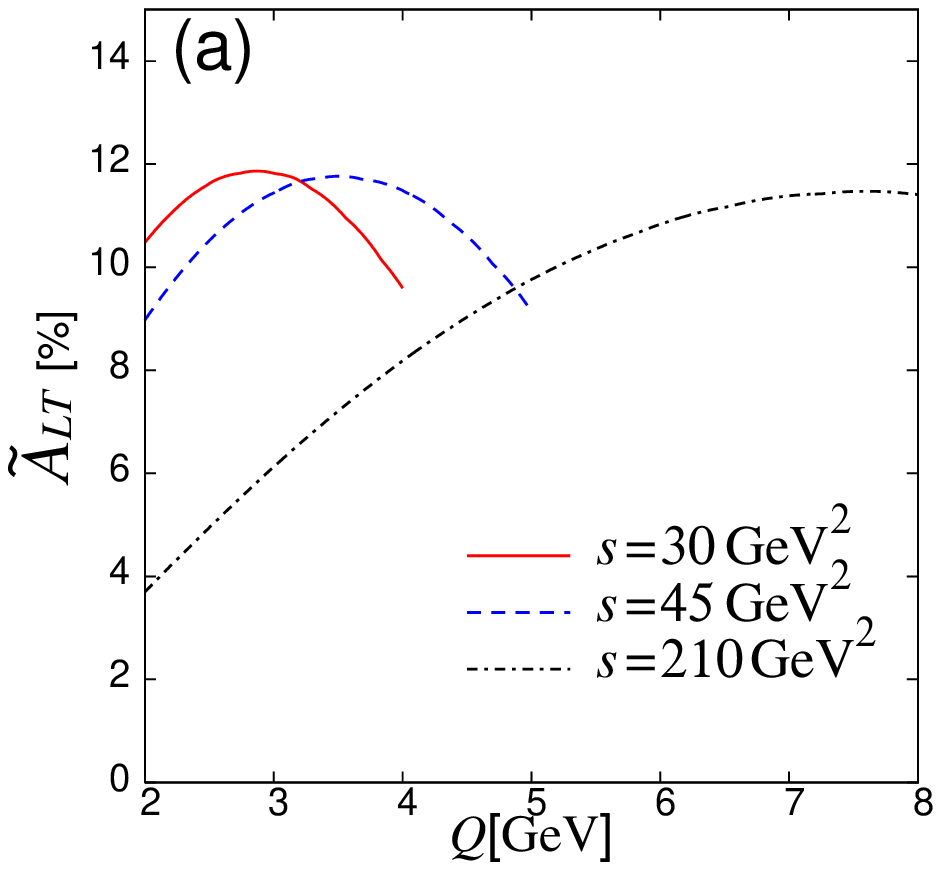}~~
\includegraphics[height=7cm]{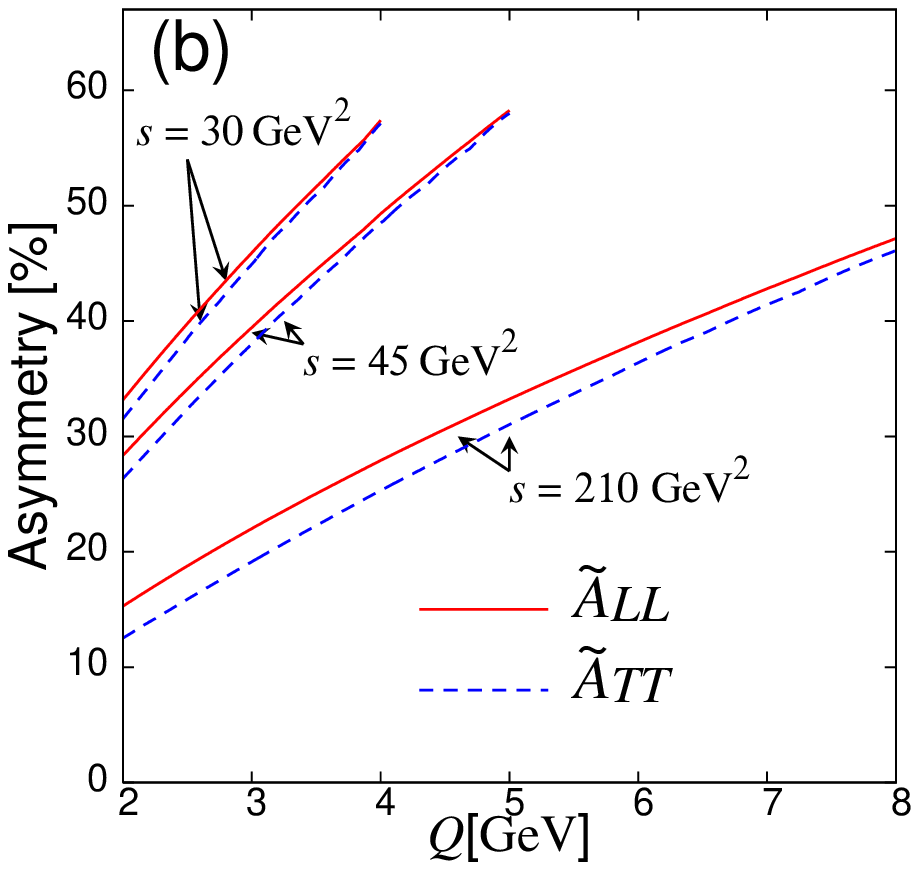}
\ec
\caption{(a)~$\widetilde{A}_{LT}$ at $x_F=0$ as a function of $Q$ 
for $s=30$, $45$ and $210$ GeV$^2$. 
(b)~$\widetilde{A}_{LL}$ and $\widetilde{A}_{TT}$ at $x_F=0$ as a function of $Q$ 
for $s=30$, $45$ and $210$ GeV$^2$. 
}
\label{fig:4}
\end{figure}

Actually, the fixed-target mode discussed above mainly probes 
the region $x_{1,2} \gtrsim 0.4$ (see (\ref{xf})), where
the transversities involved in (\ref{ATT2}), (\ref{ALT2}) are 
poorly determined at present (see the discussion above (\ref{ATT2})).
In the collider mode we probe the smaller $x_{1,2}$:
Figure~\ref{fig:3} is same as Fig.~\ref{fig:2}, but for $Q=2.5$, $4$, $6$ and $8$ GeV and 
$s=210$ GeV$^2$.
We observe the similar pattern as in Fig.~\ref{fig:2},
except that in Fig.~\ref{fig:3}(a) each of chiral-even and -odd
contributions changes its behavior between $Q=2.5$~GeV and $Q=6$~GeV.
Also, all the asymmetries become somewhat smaller for higher energy, i.e., 
for smaller $Q/\sqrt{s}$.
Actually, the mechanism relevant to this latter point leads
to the behavior commonly observed in Figs.~\ref{fig:2} and \ref{fig:3},
i.e., the increasing $\widetilde{A}_{LL}$ and $\widetilde{A}_{TT}$ for increasing $Q$, 
and the corresponding moderate increase of $\widetilde{A}_{LT}$.
The corresponding behavior is also presented in Fig.~\ref{fig:4}, 
where the relevant asymmetries 
with $x_1=x_2=Q/\sqrt{s}$ ($x_F=0$) are plotted as functions of $Q$ for
$s=30$, $45$ and $210$ GeV$^2$.
As clarified in \cite{KKT:08}, 
the $Q$ dependence of $\widetilde{A}_{LL}$ and $\widetilde{A}_{TT}$ 
shown in Fig.~\ref{fig:4}(b)
directly reflects the $x$ dependence of the corresponding 
distributions: (\ref{ATT2}) 
implies that $\widetilde{A}_{TT}$ is
controlled
by the ratio $h_{1}^u(x,Q^2)/f_{1}^u(x,Q^2)$. 
It is straightforward to see that 
the scale dependence of the $u$-quark distributions in this ratio
almost cancels between the numerator and denominator 
in the valence region relevant at GSI,
as $h_{1}^u(x,Q^2)/f_{1}^u(x,Q^2) \simeq h_{1}^u(x,1 {\rm GeV}^2)/f_{1}^u(x,1 {\rm GeV}^2)$ 
(see Fig.~3 in \cite{KKT:08}).
Thus the behavior of $h_{1}^u(x,1 {\rm GeV}^2)/f_{1}^u(x,1 {\rm GeV}^2)$ 
as a function of $x$ directly determines the $Q$-dependence of 
$\widetilde{A}_{TT}$ 
with $x=Q/\sqrt{s}$. The same logic holds for $\widetilde{A}_{LL}$.
In the present case using GRV and GRSV parameterizations,
the ratio $h_{1}^u(x,1 {\rm GeV}^2)/f_{1}^u(x,1 {\rm GeV}^2)$, as well
as $g_{1}^u(x,1 {\rm GeV}^2)/f_{1}^u(x,1 {\rm GeV}^2)$,
is actually an increasing function 
of $x$, leading to the $Q$-dependence in Fig.~\ref{fig:4}(b).
Note, this mechanism characteristic for the GSI kinematics
survives even when including the
higher order QCD corrections~\cite{KKT:08}.
For $\widetilde{A}_{LT}$, however, the cancellation 
of the scale dependence
between the numerator and denominator in (\ref{ALT2}) is less complete
due to the additional 
$y$-integral for the Wandzura-Wilczek part, which, combined with the additional
factor $x_1$ or $x_2$ ($=Q/\sqrt{s}$),
results in the novel $Q$-dependence in Fig.~\ref{fig:4}(a).
In particular, the 
suppression in the moderate $x$-region observed in Fig.~\ref{fig:1}(b)
compared with Fig.~\ref{fig:1}(a)
leads to the decreasing behavior of $\widetilde{A}_{LT}$ for increasing $Q$ in the large 
$Q$ region, 
while the increasing behavior of $\widetilde{A}_{LT}$ in the small $Q$ region is
caused by that of 
the additional factor $x_{1,2}=Q/\sqrt{s}$. 

\smallskip

To summarize, we have presented a first estimate of the
longitudinal-transverse spin asymmetry $A_{LT}$ 
for the polarized Drell-Yan process in $p\bar{p}$ collisions at GSI kinematics.
Guided by the new empirical information of
the transversity,
we performed the LO calculation of the Wandzura-Wilczek contribution to $A_{LT}$,
which is directly related to the behavior of the transversity in the valence region.
The results turned out to be significantly large, and exhibited distinguished behaviors
compared with the twist-2 asymmetries $A_{TT}$ and $A_{LL}$.
These results serve as a useful guide 
for possible future $A_{LT}$ measurement at GSI.

\smallskip

In relation to the discussion on $A_{LT}$, 
we also emphasized that the large value of $A_{TT}$ at GSI kinematics is known to be 
quite stable when including the QCD corrections, and that the behavior of $A_{TT}$
as a function of dilepton mass is controlled by the $x$-dependence of the transversity.
Thus, first of all, the measurements of $A_{TT}$ at GSI 
will provide the data that constrain the detailed shape of the transversity
in the valence region, including the large $x$ regime where our knowledge on 
transversity is poor at present.
The corresponding new information on the transversity 
will enable us to update our prediction of $A_{LT}$
in the Wandzura-Wilczek approximation.
If the strong deviation from our updated results were observed in the GSI measurements 
of $A_{LT}$,
this would provide an indication of large genuine twist-3 effect,
associated with the chiral-odd distribution $h_L$.
The QCD analysis of such data using the 
evolution equation for the corresponding twist-3 operators
will reveal the quark-gluon-quark correlation inside the nucleon.
One problem for this purpose is that
the exact form of the evolution equation governing the genuine twist-3 contributions
in $h_L$ is known to be quite sophisticated even at the LO level\,\cite{KT}.
Fortunately, 
as in the case for the similar problem in the chiral-even distribution $g_T$~\cite{ABH},
it is proved~\cite{BBKT} that, in the limit of large number of colors, $N_c \rightarrow \infty$, 
the corresponding evolution equation is simplified into the evolution of usual
DGLAP-type, with the novel anomalous dimension known in analytic form.
Since this simplification holds up to the corrections of $O(1/N_c^2) \sim 10\%$,
the large-$N_c$ evolution for the genuine twist-3 contributions in $h_L$ 
provides a powerful and practical framework 
to solve the above problem.

\section*{Acknowledgments}
We thank 
Hiroyuki Kawamura and Alexei Prokudin for valuable
discussions, and Andreas Vogt for providing us with the Fortran code of
GRV98 distribution.  
The work of K.T. was supported by the Grant-in-Aid 
for Scientific Research No. B-19340063.

\end{document}